\documentclass[a4paper,10pt,twocolumn]{article}

\usepackage[english]{babel}
\usepackage[utf8]{inputenc}
\usepackage[T1]{fontenc}

\usepackage[top=1.5cm, left=1.5cm, right=1.5cm, bottom=1.5cm]{geometry}

\renewenvironment{abstract}{\bf\small {\em\ Abstract---}}{}

\usepackage{amsmath,amsfonts,amssymb,amsthm}
\usepackage{graphicx}
\usepackage{xspace}
\usepackage{tikz}
\usepackage{epstopdf}
\usepackage{subcaption}
\newfont{\ninept}{ptmr at 9pt}

\usepackage{hyperref}
\usepackage{mathtools}

\usepackage{amsfonts,amssymb,amsmath,amsthm}

\newcommand{\Amap}{\bs {\sf A}}

\newcommand{\supp}{{\rm supp}\,}
\newcommand{\tinv}[1]{{\textstyle\frac{1}{#1}}}

\newcommand{\bs}{\boldsymbol}
\newcommand{\bb}{\mathbb}
\newcommand{\cl}{\mathcal}

\newcommand{\ts}{\textstyle}

\newcommand{\ie}{\emph{i.e.}, }
\newcommand{\eg}{\emph{e.g.}, }

\newcommand{\sq}{\vspace{-1.5mm}}
\newcommand{\hsq}{\vspace{-0.5mm}}
\newcommand{\im}{{\sf i}\mkern1mu} %

\usepackage{amsmath,amsfonts,amssymb,amsthm}
\usepackage{graphicx}
\usepackage{xspace}
\usepackage{tikz}
\usepackage{epstopdf}
\usepackage{subcaption}

\usepackage{hyperref}
\usepackage{mathtools}

\DeclareMathOperator{\diag}{diag}

\makeatletter
\newcommand{\RemoveAlgoNumber}{\renewcommand{\fnum@algocf}{\AlCapSty{\AlCapFnt\algorithmcfname}}}
\newcommand{\RevertAlgoNumber}{\algocf@resetfnum}
\makeatother

\newcommand{\iid}{%
    \ifmmode
        \mathrm{i.i.d.}%
    \else%
        i.i.d.\@\xspace%
    \fi%
}

\makeatletter
\renewcommand{\section}{\@startsection {section}{1}{\z@}%
{-3.5ex \@plus -1ex \@minus -.2ex}%
{2.3ex \@plus.2ex}%
{\normalfont\large\bfseries}}
\makeatother 

\title{\vspace{-0.5cm}An extreme bit-rate reduction scheme for 2D radar localization}

\author{Thomas Feuillen$^1$, Luc Vandendorpe$^1$ and Laurent Jacques$^{2}$.\\
  \footnotesize $^1$CoSy, $^2$ISPGroup, ICTEAM/ELEN, UCLouvain, Louvain-la-Neuve, Belgium.  \thanks{LJ is funded by the F.R.S.-FNRS. Part of this study is funded by the project {\sc AlterSense} (MIS-FNRS).}} \date{\empty}

\begin{document}
\maketitle
\begin{abstract}
In this paper, we further expand on the work in \cite{COSERA} that focused on the localization of targets in a 2D space using 1-bit dithered measurements coming from a 2 receiving antennae radar. Our aim is to further reduce the hardware requirements and bit rate, by dropping one of the baseband IQ channel from each receiving antenna. To that end, the structure of the received signals is exploited to recover the positions of multiple targets. Simulations are performed to highlight the accuracy and limitations of the proposed scheme under severe bit-rate reduction.
\end{abstract}
\vspace{-0.55cm}
\section{Introduction}
\vspace{-0.17cm}
In radar processing, \textit{Compressive Sensing} (CS) offers the potential to simplify the acquisition process \cite{bara} or to use super resolution algorithms to solve ambiguous estimation problems \cite{strohmer}. However, the underlying assumption of such schemes is the availability of high resolution radar signals, requiring high bit-rate data transmission to a processing unit. 

In this article, we aim to break this assumption and to further explore the reconstruction of the target scene on the basis of radar signals acquired under a harsh bit-rate acquisition process, \ie a regime where classic estimation methods fail (\eg Maximum Likelihood~\cite{mkay}).
 
We propose to reconstruct the target scene in the extreme 1-bit measurement regime, in a similar way to only recording the sign of each sample~\cite{BB2008, JLBB2013, PV2013}. We further reduce the information by only sampling either the real (I channel) or imaginary (Q channel) part of the signal. This has interesting implications in terms of bit-rate as well as production cost reduction in radar modules.
We show that estimating the 2D-localization of multiple targets observed from a radar system with two antennae under the harsh bit-rate requirement and with half of the channels recorded is feasible. 

We further reveal, through Monte Carlo simulations, a certain trade-off, for a fixed bit-rate, between the number of measurements and the resolution by comparing the performances of \textit{Projected Back Projection} (PBP) under multiple scenarios involving one or two targets and different measurement numbers and resolutions. The limitations induced by the channel dropping model are highlighted for both the quantized and unquantized schemes.

The rest of this paper is structured as follows. The radar signal model is introduced in Sec.~\ref{radar_model}. The bit-rate reduction model is presented in Sec.~\ref{channel} and Sec.~\ref{QCS}. The PBP algorithm is introduced in Sec.~\ref{PBP}. In Sec.~\ref{results}, the proposed scheme is tested under different scenarii using Monte Carlo simulations before the conclusion. 
\vspace{-0.55cm}
\section{Radar system model}
\vspace{-0.17cm}
\label{radar_model}
\begin{figure}[!t]
\centering
\includegraphics[width=0.4\columnwidth]{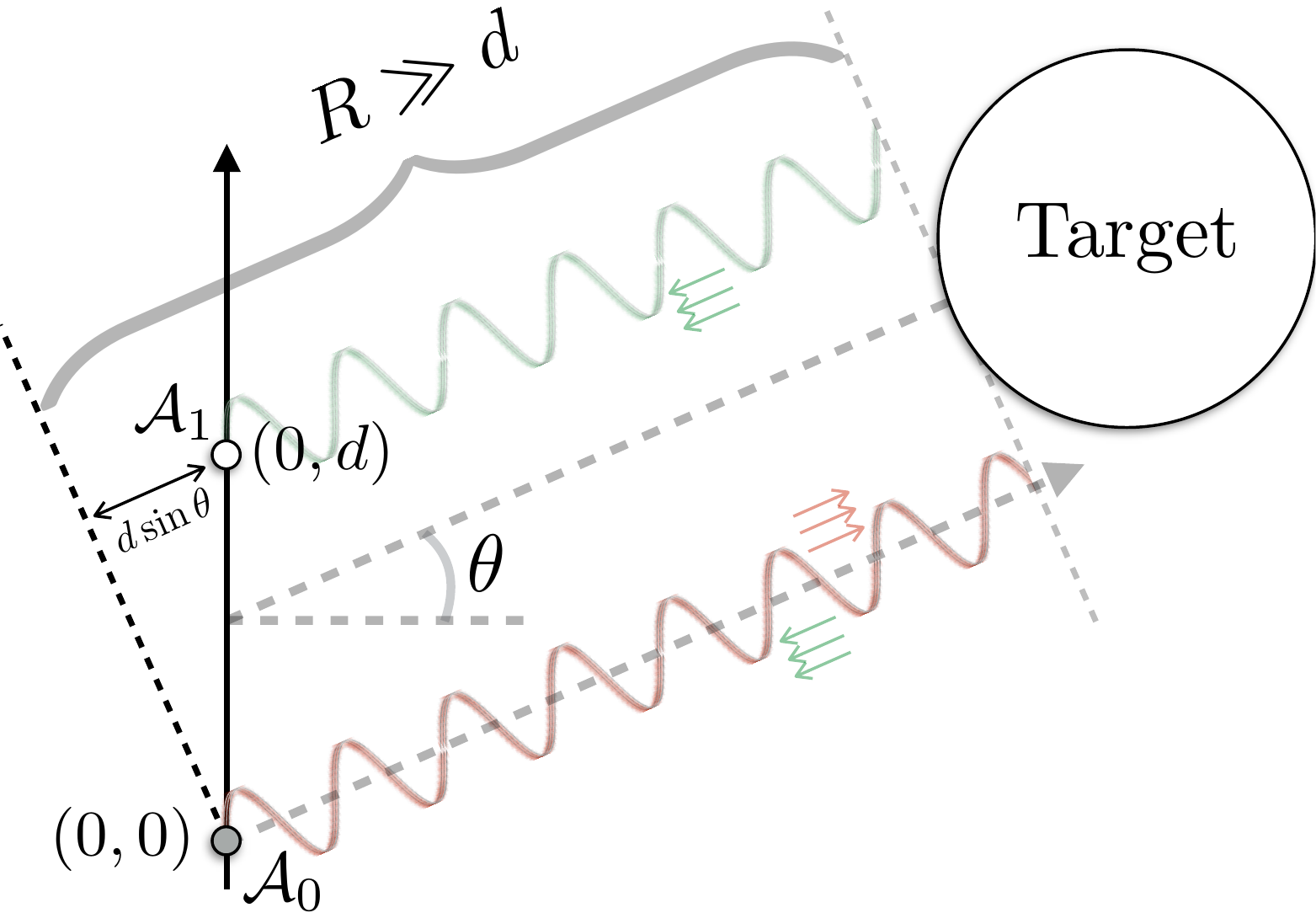}
\caption{Illustration of the two antennas radar system.\sq\sq\sq}
\label{ULA}
\end{figure}
This paper considers a \textit{Frequency Modulated Continuous Wave} (FMCW) radar. The system model is identical to the one used in \cite{COSERA}. A frequency modulated signal is transmitted. After coherent demodulation of the received radar signals, the acquisition model links the sampling time with the frequency that is being transmitted. For a regular sampling at rate $M/T_r$, $T_r$ being the duration of one ramp, the sampled frequencies are $f_m := f_0 +B\frac{m}{M}$, $1\leq m \leq M$, so that, at the $m^{\rm th}$ frequency and the $p^{th}$ antenna, the following sample is received:\hsq
\begin{align}
    \label{approxf0}
   \ts \Gamma_{mp}&= \ts x\, e^{-\im 2 \pi f_m \tau_p} \approx x\, e^{-\im 2 \pi f_m  \frac{2 R}{c}} e^{-\im 2 \pi f_0\frac{p d \sin \theta}{c}},\hsq
\end{align}
where $\tau_p=\frac{2R}{c}+\frac{d \sin \theta}{c}$ is the delay from the range $R$ and angle $\theta$ (see Fig. \ref{ULA}), $x$ is the received amplitude after the coherent demodulation. The approximation in \eqref{approxf0} is reasonable for K-band radars where the bandwidth $B \ll f_0$, \ie ${B=250{\rm MHz}}$ and ${f_0=24{\rm GHz}}$ respectively.

For a scene with $K$ targets, recasting \eqref{approxf0} into a linear matrix sensing model and taking advantages of the phase relation between the two antennae, the sensed signals $\bs \Gamma = \{\Gamma_{mp}\}_{mp} \in \bb C^{M \times 2}$ are\hsq
\begin{equation}
\label{eq:setting-signal}
\ts \bs \Gamma = [\bs \gamma_1, \bs \gamma_2] = \bs \Phi\, [\bs x, \bs G  \bs x],\hsq
\end{equation}
where $\bs x \in \mathbb{C}^{N}$ encodes the range profile, \ie $x_n \neq 0$ if there exists a target at range $R_n$, ${\|\bs x\|_0 := |\supp \bs x| \le K \ll N}$, ${\bs \Phi = \{ e^{-\im \frac{4\pi}{c} f_m R_n }\}_{mn} \in \bb C ^{M\times N}}$ is the \emph{range} measurement matrix, $\bs G =\diag(e^{-\im \frac{2 \pi}{c} f_0 d \sin \theta_1},\,\cdots, e^{-\im \frac{2 \pi}{c} f_0 d \sin \theta_N})$ with $\supp (\bs \theta)= \supp (\bs x)$,\ie $\bs G$ is the phase difference between the first and second receiving antennae. Therefore, the 2D-localization problem is tantamount to estimating the support $T$ of $\bs x$ from the sensing model \eqref{eq:setting-signal}. Comparing the phases of $\bs x$ and $\bs G \bs x$ on the index set $T$ then enables to deduce the angles $\bs \theta_T$. The frequencies $f_m$ are chosen in a \textit{semi-random} fashion, \ie for $M$ measurements, $ \lfloor \frac{M}{N}\rfloor$ complete ramps are sampled and $M \bmod N$ frequencies are uniformly sampled at random on the last ramp. Note that, in the absence of quantization, inverting \eqref{eq:setting-signal} can be solved using Maximum Likelihood \cite{mkay} if $M \geq N$, or using CS algorithms (\eg IHT \cite{BD2009} or CoSaMP \cite{NT2009}) if $M \leq N$. 
\vspace{-0.55cm}
\section{Channel dropping model}
\vspace{-0.17cm}
\label{channel}
The coherent demodulation of the signal received by the antennae produces complex signals that are each transmitted on 2 channels representing the real and the imaginary parts. Common radar acquisition scheme requires the sampling of both of these channels, resulting in $2M$ measurements per antenna. We propose to acquire only half of the channels as follows:
\begin{equation}
  \bs y_1=\text{Re}\{\bs \gamma_1  \},\quad \bs y_2=\im \text{Im} \{\bs \gamma_2  \}
  \label{eq:setting-signal2} 
\end{equation}
This simplification, however, comes at a cost. As explained in \cite{COSERA}, range estimation is equivalent to estimating the support of the spectrum of the received signals.  $\bs y_1$ and $-\im \bs y_2$, being purely real signals, their spectrum are by definition symmetric, doubling the sparsity in an ambiguous way. To partly solve this problem we observe that for one target at ($R_k$, $\theta$)  : 
\begin{align}
\label{yhat}
    \hat{\bs y}&=\bs y_1+\bs y_2\\
    &=\left(1+G_{kk}\right)e^{-\im 2 \pi f_m \frac{2R}{c}}+\left(1-G_{kk}^*\right) e^{\im 2\pi f_m \frac{2R}{c}}
\end{align}
Eq.\eqref{yhat} shows, provided $G$ tends to $1$ (\ie for small angles) that the unambiguous support of the complex signal can be approximated (\ie $ \supp(\Phi^*\hat{ \bs y}) \approx \supp(\bs x_1)$). Given the Fourier nature of \eqref{eq:setting-signal} some ranges will remain ambiguous (\eg for ranges $R=0$ and $R=R_{max}/2$ which corresponds to the \textit{DC} and maximum spectral components). It is important to note that this limitation affects the signal regardless of the quantization.
\vspace{-0.55cm}
\section{Quantization model}
\vspace{-0.17cm}
\label{QCS}
In this work, we propose to quantize the observations $\bs \Gamma$ achieved in the digital beamforming model \eqref{eq:setting-signal2}. Our quantization procedure relies on a uniform scalar quantizer $\lambda \in \bb R \mapsto \cl Q(\lambda) = \delta \lfloor \frac{\lambda}{\delta}\rfloor + \frac{\delta}{2} \in \bb Z_\delta := \delta \bb Z + \frac{\delta}{2}$, with quantization width $\delta > 0$ applied entrywise onto vector or matrices, and separately onto the real and imaginary parts if these objects are complex. Our global objective is thus to estimate the localization of targets, as encoded in the matrix $\bs X = (\bs x_1, \bs x_2) = (\bs x, \bs G \bs x) \in  \bb C^{N\times 2}$, from the quantized observation model\hsq
\begin{equation}
  \label{eq:QRadar-system}
\bs Z = \Amap^D(\bs X) := \big(\text{Re}\{\Amap(\bs x_1)\}, \im\text{Im}\{\Amap(\bs x_2)\} \big),\hsq
\end{equation}
with $\bs u \in \bb C^N \mapsto \Amap(\bs u) = \cl Q(\bs \Phi \bs u + \bs \xi) \in \bb Z^M_\delta + \im \bb Z^M_{\delta}$. In $\Amap$, a uniform random \textit{dithering} $\bs \xi \in \bb C^{M}$, \ie $\xi_{m}  \sim_{\iid} \cl U{([-\frac{\delta}{2},\frac{\delta}{2}])}+\im U{([-\frac{\delta}{2},\frac{\delta}{2}])} $ for all $m \in [M]$, is added to the quantizer input. For real sensing models, such a dithering attenuates the impact of the quantizer on the estimation of sparse/compressible signals in quantized CS \cite{PB2012,JC2017,XJ2018}.
\vspace{-0.55cm}
\section{2D Target Localization in Quantized Radar}
\vspace{-0.17cm}
\label{PBP}
We adopt here a simple method, which is an adaptation of PBP proposed in \cite{XJ2018}. The estimate is, first, defined from the backprojection:
\begin{equation}
  \label{eq:PBP}
  \hat{\bs X} = \tinv{M} \bs \Phi^* \bs Z.
\end{equation}
Next, to recover the support $\hat{T}$ from $\hat{\bs X}$ we rely on the approximation defined in \eqref{yhat} to partly cancel the symmetrical shape of $\hat{\bs{x}}_1$ and $\hat{\bs{x}}_2$ : 
\begin{equation}
    \hat{T}=\supp \left( \mathcal{H}_K^\text{Sym}(\hat{\bs{x}}_1+\hat{\bs{x}}_2) \right)
    \label{T}
\end{equation}
where $\mathcal{H}_K^\text{Sym}(\cdot)$ is the hard thresholding operator which takes the $K$ biggest elements excluding the weakest symmetrical elements. The targets are localized in the polar coordinates $(R_n,\theta_n)$ for all $n \in \hat T$, with $\theta_n = \arcsin\big(\frac{c}{2\pi f_0 d} \angle(\hat x_2[n]^* \hat x_1[n])\big)$. It is worth mentioning that $\mathcal{H}_K^\text{Sym}(\cdot)$ imposes new limitations on the relative position of targets when $K\ge 2$.
\vspace{-0.55cm}
\section{Results}
\begin{figure}[!t]
    \centering
    \begin{subfigure}[b]{0.10\textwidth}
        \includegraphics[height=2.8cm]{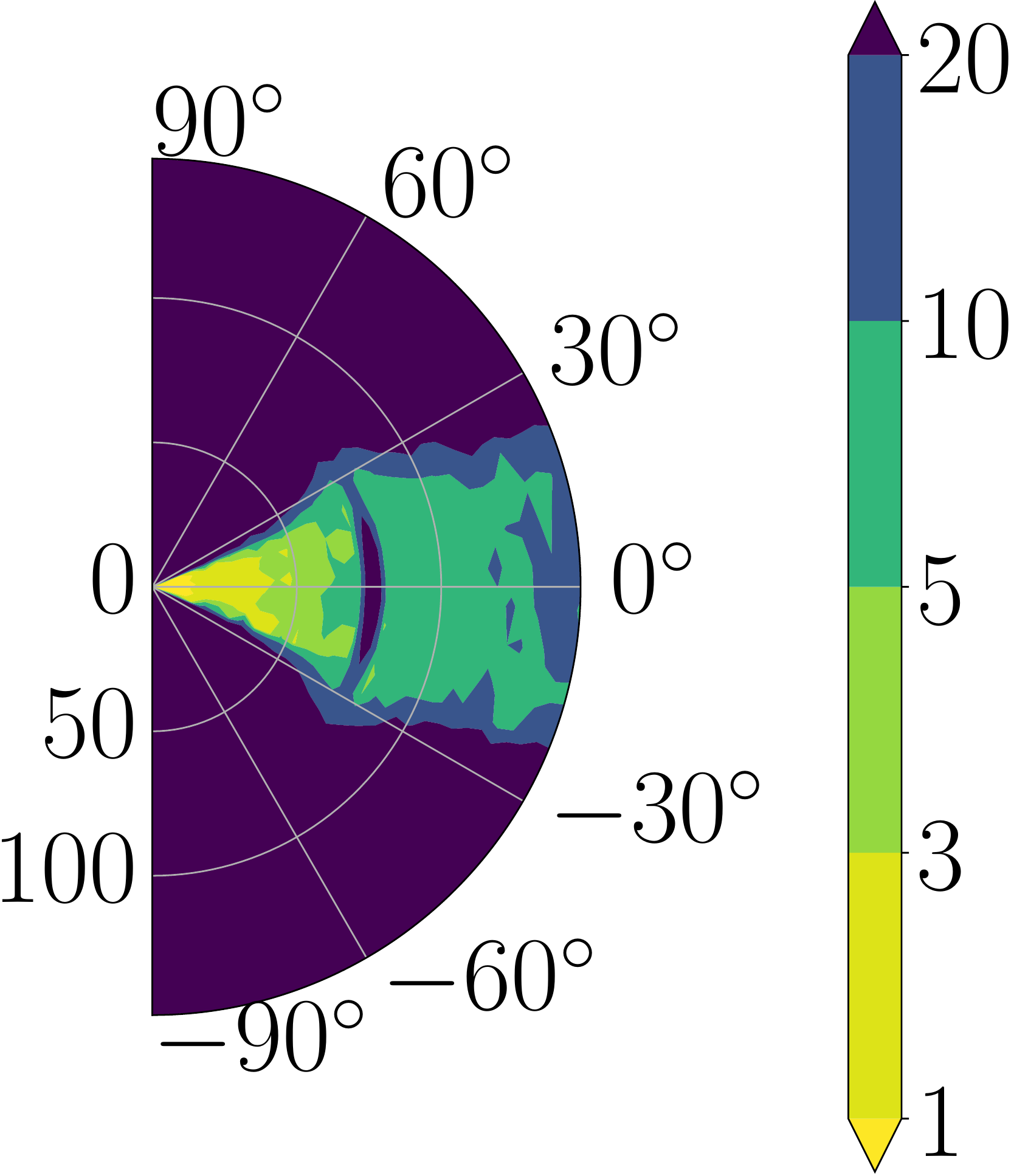}
        \caption{}
        \label{fig:11ND}
    \end{subfigure}
    \begin{subfigure}[b]{0.10\textwidth}
        \includegraphics[height=2.8cm]{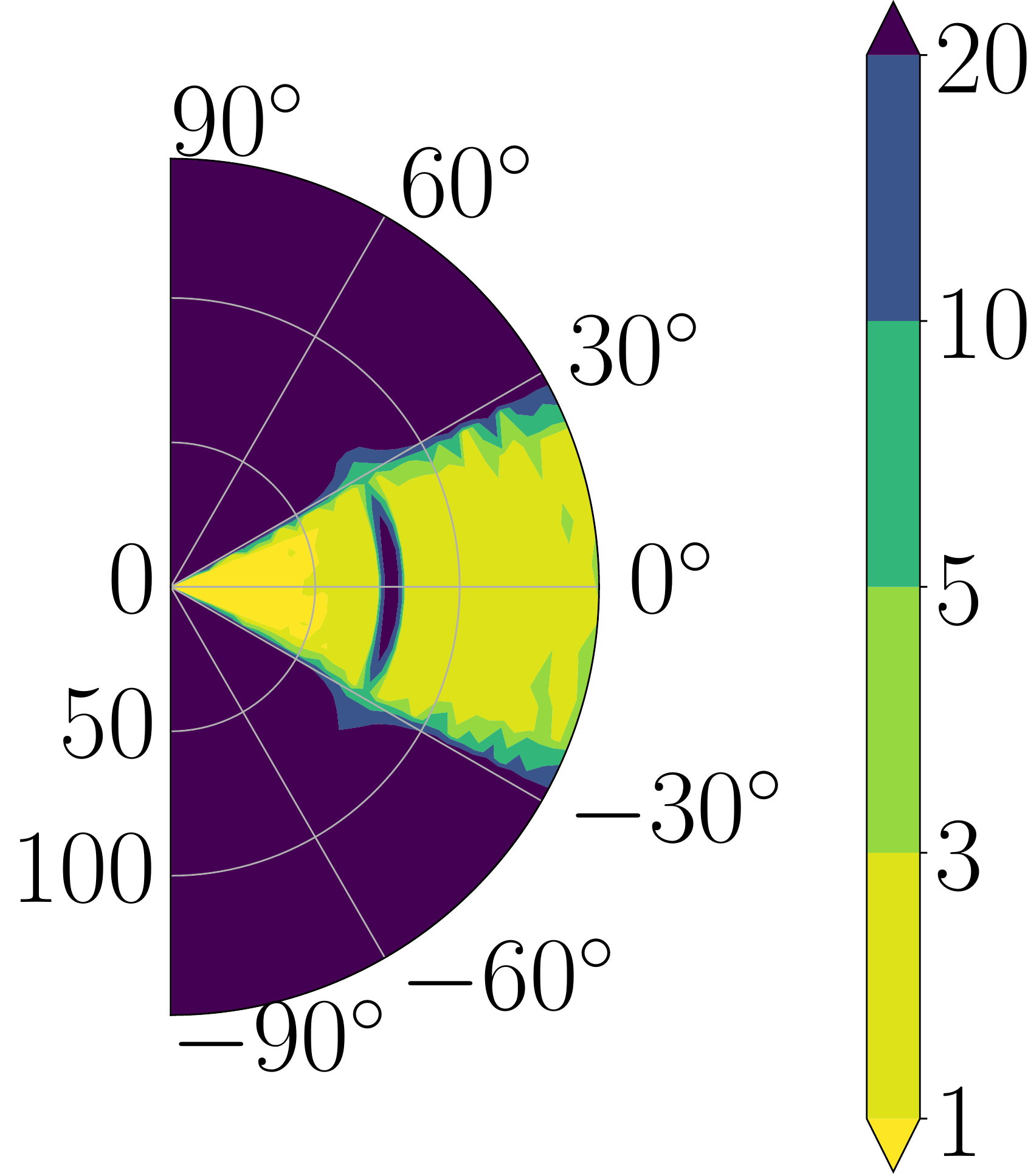}
        \caption{}
        \label{fig:12ND}
    \end{subfigure}
    \begin{subfigure}[b]{0.10\textwidth}
        \includegraphics[height=2.8cm]{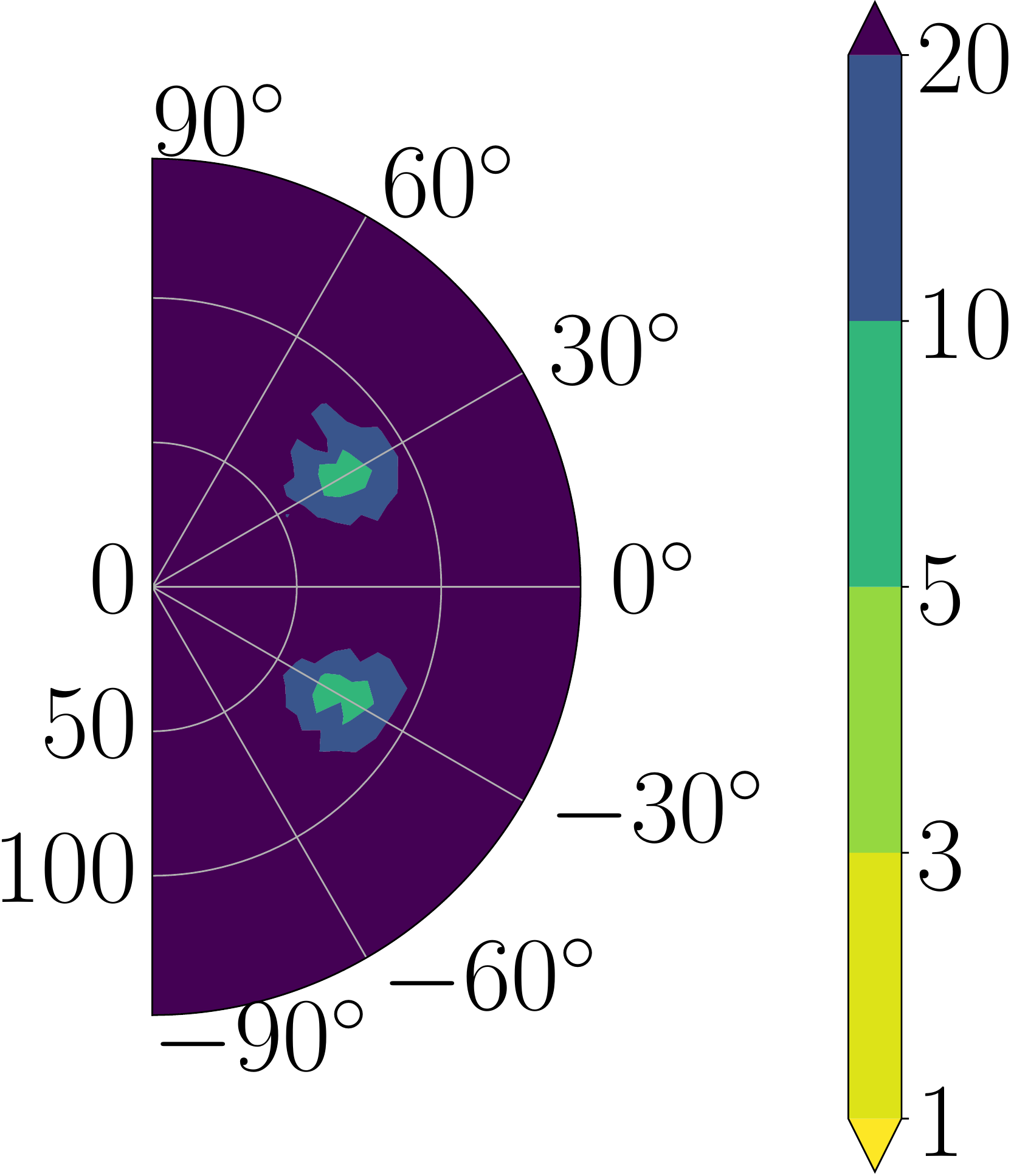}
        \caption{}
        \label{fig:11D}
    \end{subfigure}
    \begin{subfigure}[b]{0.12\textwidth}
        \includegraphics[height=2.8cm]{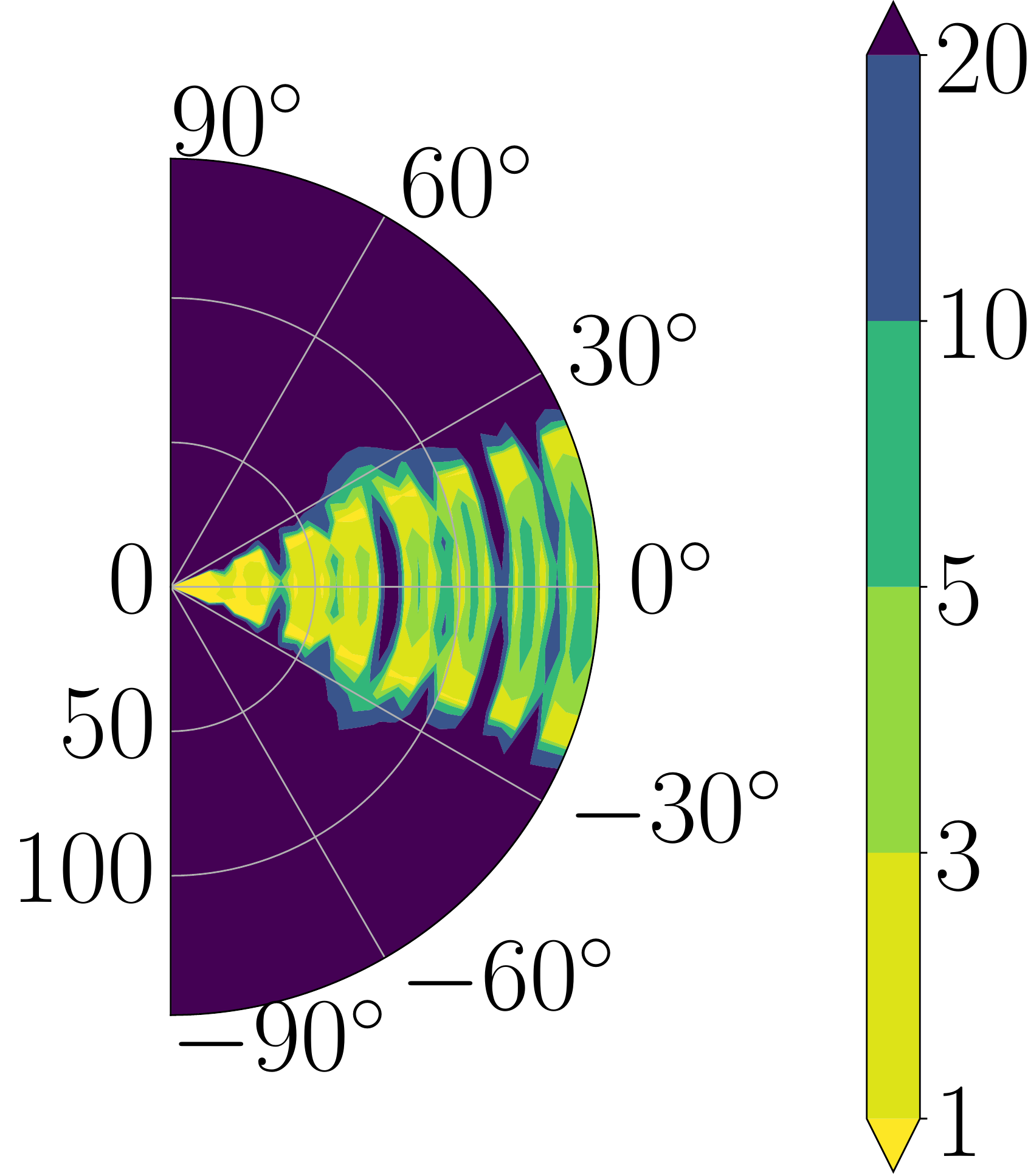}
        \caption{}
        \label{fig:12D}
    \end{subfigure}\vspace{-4mm}
  \caption{\ninept Positions error in meters for Monte Carlo simulations with one target; (a) 1 bit dithered with $\frac{M}{N}=20\%$ ; (b) 1 bit dithered with $\frac{M}{N}=200\%$; (c)  1 bit dithered with $\frac{M}{N}=200\%$ using PBP in \cite{COSERA}; (d)  1 bit non-dithered with $\frac{M}{N}=200\%$, respectively.\sq}
  \label{fig:K1_sep}
\end{figure}

\begin{figure}[!t]
    \centering
    \begin{subfigure}[b]{0.10\textwidth}
        \includegraphics[height=2.8cm]{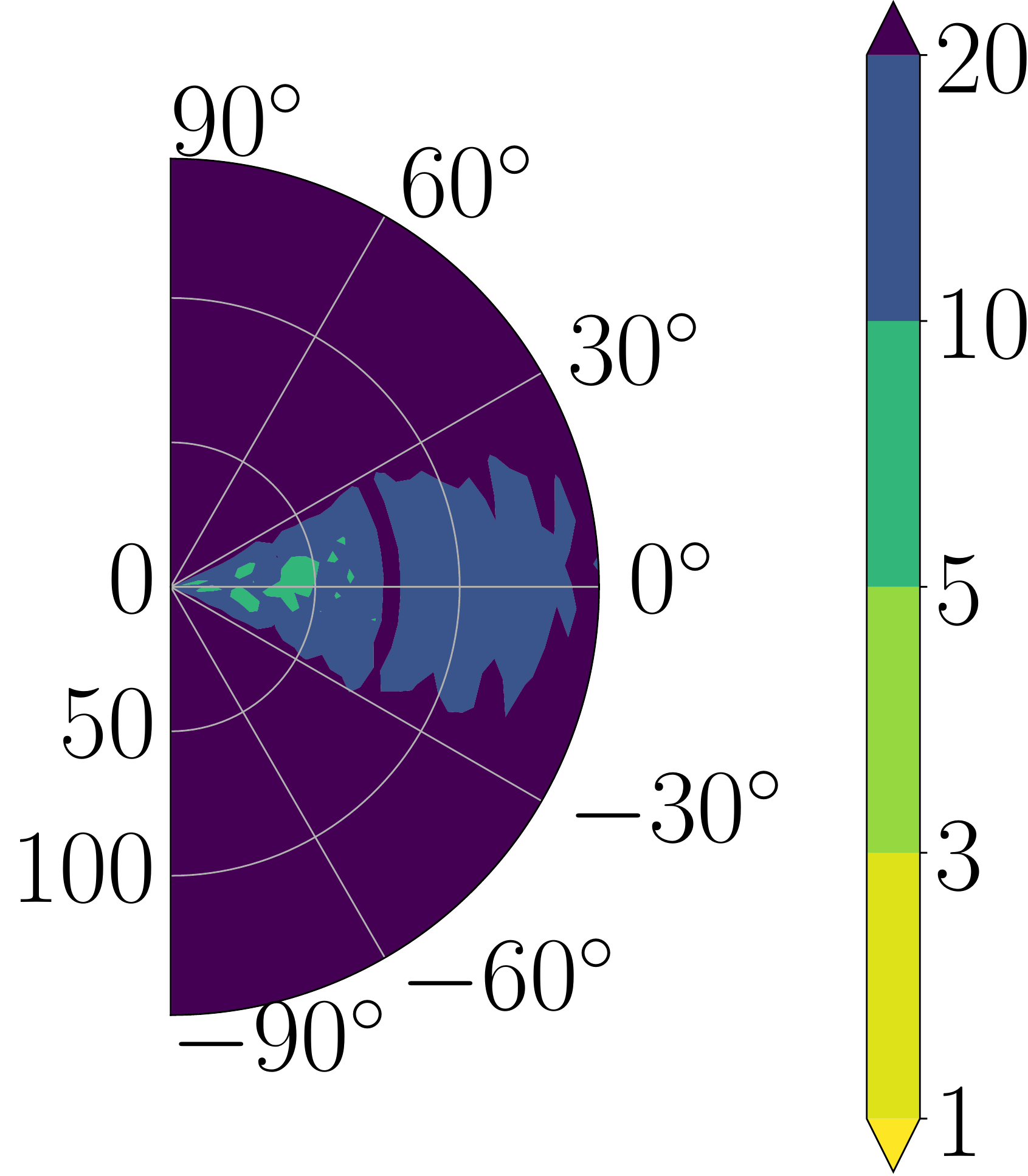}
        \caption{}
        \label{2fig:11ND}
    \end{subfigure}
    \begin{subfigure}[b]{0.10\textwidth}
        \includegraphics[height=2.8cm]{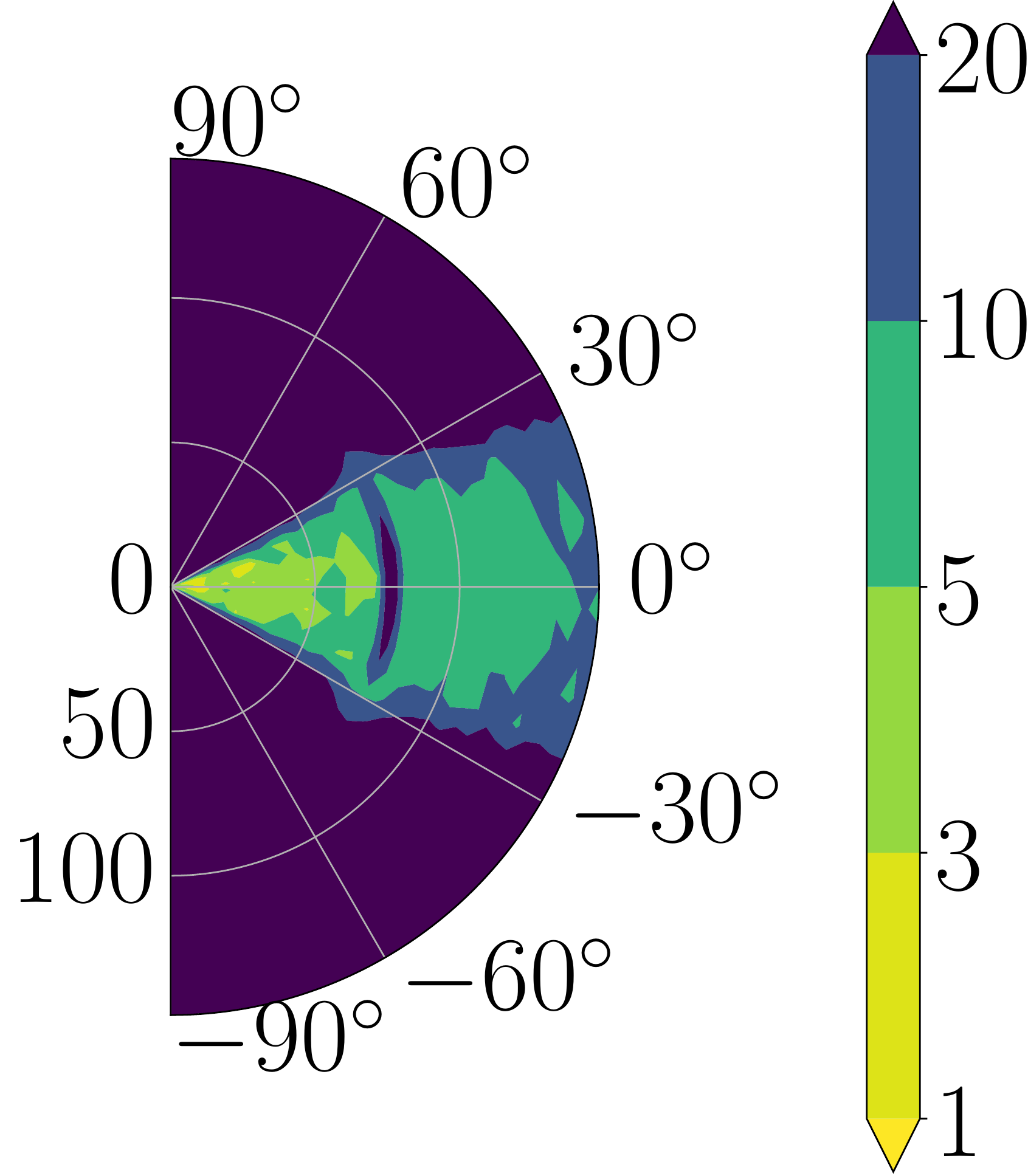}
        \caption{}
        \label{2fig:12ND}
    \end{subfigure}
    \begin{subfigure}[b]{0.10\textwidth}
        \includegraphics[height=2.8cm]{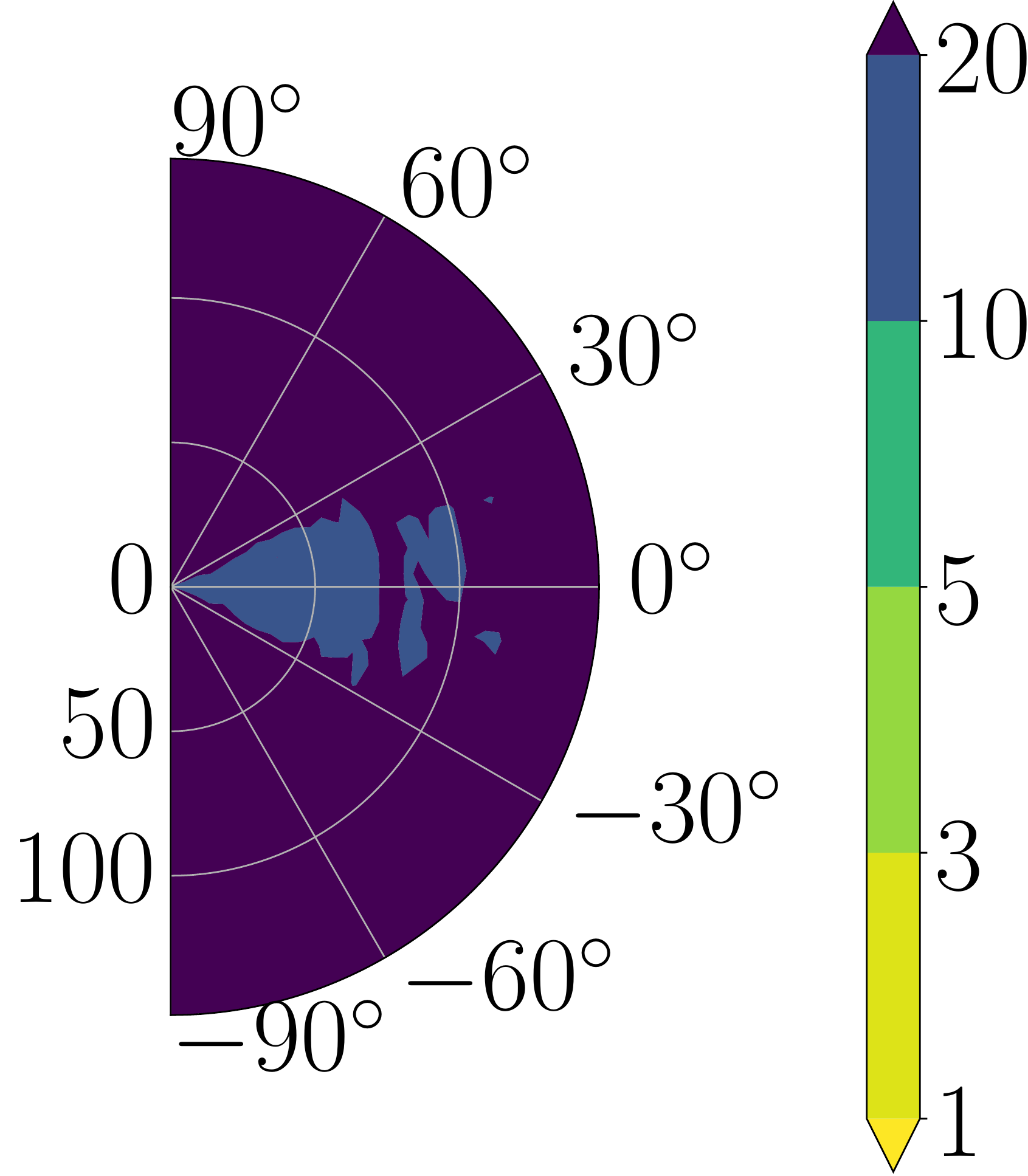}
        \caption{}
        \label{2fig:11D}
    \end{subfigure}
    \begin{subfigure}[b]{0.12\textwidth}
        \includegraphics[height=2.8cm]{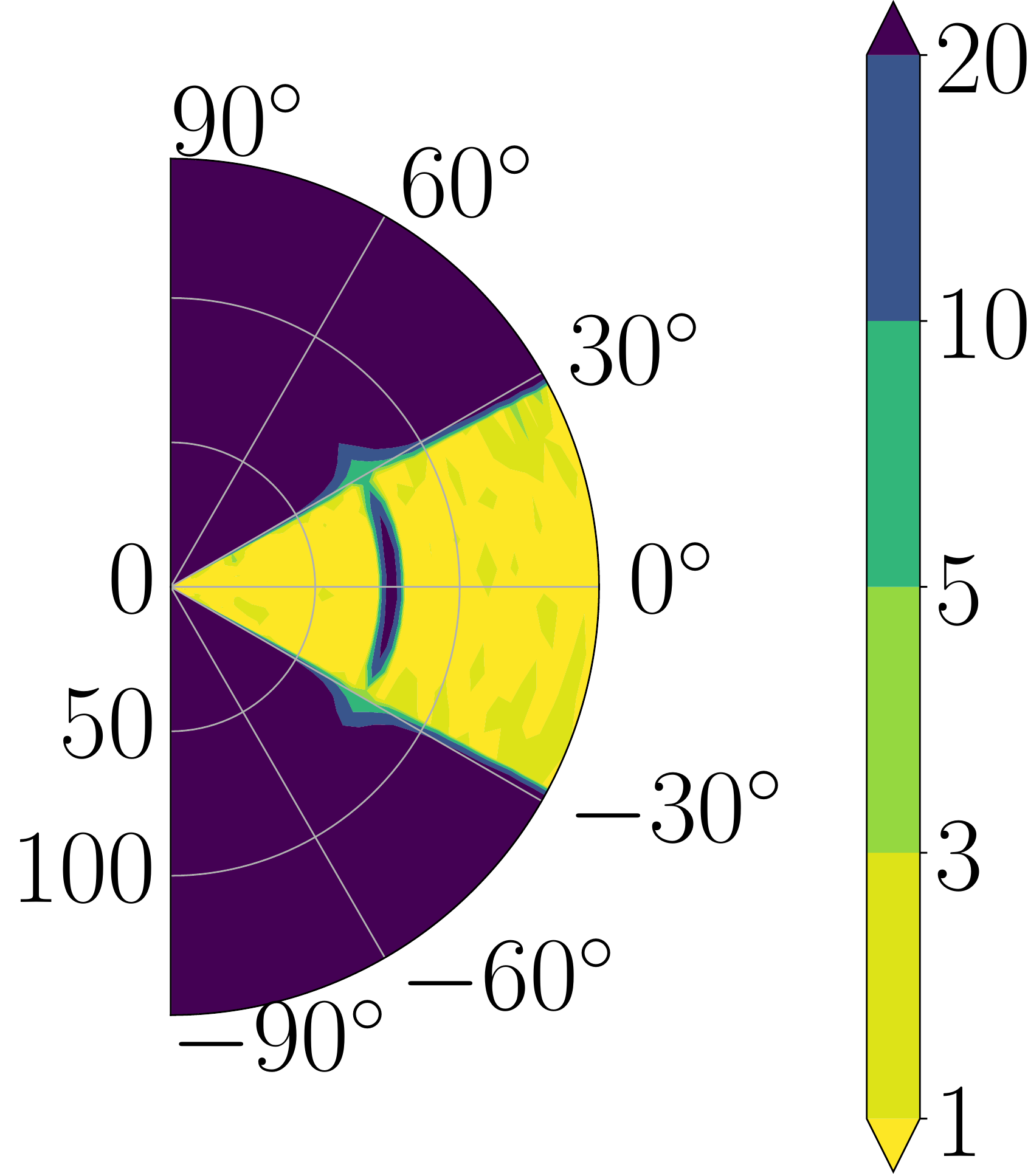}
        \caption{}
        \label{2fig:12D}
    \end{subfigure}\vspace{-4mm}
  \caption{\ninept Positions error in meters for Monte Carlo simulations with two targets; (a) 1 bit non-dithered with $\frac{M}{N}=200\%$ ; (b) 1 bit dithered with $\frac{M}{N}=200\%$; (c)  32 bit non-dithered with $\frac{M}{N}=6.25\%$; (d) 32 bit non-dithered with $\frac{M}{N}=100\%$, respectively.\sq}
  \label{fig:K2_sep}
\end{figure}

\vspace{-0.17cm}
\label{results}
We simulate the working mode of a noiseless K-Band radar, \ie giving $f_0 = 24{\rm GHz}$ and a bandwidth of $B=250 {\rm MHz}$. In all our simulations, we set the number of ranges $N$ to $256$, giving a range limit of $R_{\max}=153.6$m and a range resolution of $0.6$m. Targets' localization are picked uniformly at random in a $40\times 30$ discretized polar domain $(R, \theta) \in [0,R_{\max}] \times [-\pi/2, \pi/2]$. The quality of the position estimation is simply measured as $\min_k |R e^{\im \theta}-\hat R_k e^{\im \hat \theta_k}|$, \ie the distance between the true target location and the closest estimated targets in $(\hat{ \bs R},\hat{ \bs \theta})$. This quality measure is then averaged over runs which have the same position $(R,\theta)$. These results are reported in a 2D polar graph (Fig.~\ref{fig:K1_sep}, Fig.~\ref{fig:K2_sep}). A more detailed explanation of the simulation set-up can be found in \cite{COSERA}.
Fig.\,\ref{fig:K1_sep} shows the performances of the proposed scheme for different configurations for $K=1$. Fig.\,\ref{fig:11ND} and Fig.\,\ref{fig:12ND} show that the localization error decreases as $M$ increases. Furthermore the bit-rate reduction compared to classic sampling scheme is of $99.69\%$ and $96.87\%$ respectively. Fig.\,\ref{fig:11D} is the performance obtained when the scheme in \cite{COSERA} is applied after channel dropping. The hard thresholding operator introduced in Eq.\eqref{T} is shown to be the key to recover from the omissions of channels. The artifacts showcased in \cite{COSERA} resulting from the 1 bit non dithered quantization is still present in Fig.\,\ref{fig:12D}. The maximum angle that can be recovered is also reduced from \cite{COSERA} which is consistent with the approximation \eqref{yhat}.
The sparsity is increased to $K=2$ in Fig.\,\ref{fig:K2_sep}. Similarly to what has been observed in \cite{COSERA}, the dithered scheme in Fig.\,\ref{2fig:12ND} clearly outperforms the non-dithered one in Fig.\,\ref{2fig:11ND}. Fig.\,\ref{2fig:11D} shows the performances of the scheme when constrained to the same bit rate, thus with a smaller $M$, with a classic full acquisition scheme. The 1 bit dithered scheme represents an interesting trade off between the bit resolution and the transmitted bit-rate when compared to Fig.\,\ref{2fig:12D}. Finally the ambiguity in $\frac{R_{max}}{2}$ is present, as expected in every presented schemes.   
\vspace{-0.55cm}
\section{Conclusion}
\vspace{-0.17cm}
\label{sec:conclusion}
In this work, we have studied the localization of multiple targets configurations by using two receiving antennae combined with 1-bit radar quantization, which resulted into the QCS model \eqref{eq:QRadar-system}. We showed that the dithered scheme exhibits interesting performances given the harsh bit rate reduction and the channel dropping model. When compared to the other schemes, it was shown to be the best trade off between accuracy and bit-rate.

\end{document}